\documentclass[aip,pop,12pt,reprint]{revtex4-1}

\usepackage{graphicx}
\usepackage{amsmath}
\usepackage{amssymb}
\usepackage{url}
\usepackage{color}
\usepackage{esint}
\usepackage{float}
\usepackage{calc}
\usepackage{ae}
\usepackage{epstopdf}
\usepackage{epsfig}

\renewcommand{\b}[1]{\ensuremath{\mathbf{#1}}}
\newcommand{\be}{\begin{equation}}
\newcommand{\ee}{\end{equation}}
\newcommand{\bea}{\begin{eqnarray}}
\newcommand{\eea}{\end{eqnarray}}

\newcommand{\rd}{\ensuremath{\mathrm{d}}}
\newcommand{\sub}[1]{\ensuremath{_{\text{#1}}}}

\newcommand{\chalmers}{Department of Applied Physics, Chalmers University of Technology, Gothenburg, Sweden}

\newcommand{\Eq}[1]{Eq.~\eqref{#1}}
\newcommand{\Fig}[1]{Fig.~\ref{#1}}

\newcommand{\zeff}{\ensuremath{Z_{\text{eff}}}}

\begin{document}


\title{Effect of bremsstrahlung radiation emission on distributions of runaway electrons in magnetized plasmas}

\author{O. Embr\'eus}
\affiliation{\chalmers}
\author{A. Stahl}
\affiliation{\chalmers}
\author{S. Newton}
\affiliation{\chalmers}
\affiliation{CCFE, Culham Science Centre, Abingdon, Oxon, OX14 3DB, United Kingdom}
\author{G. Papp}
\affiliation{Max-Planck Institute for Plasma Physics, D-85748 Garching, Germany}
\author{E.~Hirvijoki}
\affiliation{\chalmers}
\author{T. F\"{u}l\"{o}p}
\affiliation{\chalmers}

\date{\today}

\begin{abstract}
  Bremsstrahlung radiation is an important energy loss mechanism for
  energetic electrons in plasmas.  In this paper we investigate the
  effect of bremsstrahlung radiation reaction on the electron distribution in 2D momentum
  space. We show that the emission of bremsstrahlung radiation leads to
  non-monotonic features in the electron distribution function and
  describe how the simultaneous inclusion of synchrotron and
  bremsstrahlung radiation losses affects the dynamics of fast
  electrons.  We give quantitative expressions for (1) the maximum
  electron energy attainable in the presence of bremsstrahlung losses and (2)
  when bremsstrahlung radiation losses are expected to have a stronger
  effect than synchrotron losses, and verify these expressions
  numerically. We find that, in typical tokamak scenarios, synchrotron
  radiation losses will dominate over bremsstrahlung losses, except in cases 
  of very high density, such as during massive gas injection.
\end{abstract}

\maketitle


\section{Introduction}
\label{sec:intro}

Energetic electrons are ubiquitous in plasmas, and bremsstrahlung
radiation is one of their most important energy loss mechanisms
\cite{heitler1954}.  At sufficiently high energy, around a few hundred
megaelectronvolts in hydrogen plasmas, the loss associated with the emission of
bremsstrahlung radiation dominates the energy lost by the fast electron through elastic
collisions. Bremsstrahlung can also strongly affect electrons at
somewhat lower energies, around a few tens of megaelectronvolts,
particularly in high-density plasmas where the magnetic field is not
too strong.

An important electron acceleration process, producing energetic
electrons in both space and laboratory plasmas, is the runaway
mechanism \cite{dreicer1959}.
This requires an electric field of intermediate strength, parallel to
the magnetic field in a magnetized plasma: it must be stronger than a
critical field to overcome the friction force that fast electrons
experience due to collisions \cite{connorhastie1975}, while
sufficiently weaker than the Dreicer field, above which the entire
electron population will promptly be accelerated \cite{dreicer1959}.
A fraction of the charged particles can then detach from the bulk
population and accelerate to high energy, where radiative losses become
important.

Disruptions in fusion devices can induce strong, time-dependent
electric fields, leading to the generation of runaway electrons (REs)
with energies reaching several tens of megaelectronvolts -- it has
been proposed that runaway electrons in ITER \cite{ITER} would be able
to reach energies on the order of 100 MeV \cite{papp2011}.  These
electrons could severely damage plasma facing components. A key issue
for ITER and other future tokamaks is therefore to increase the
understanding of the dynamics of these relativistic electrons, with
the specific goal of reducing their energy and destructive potential
\cite{hollmann_ITER_DMS}.  Previous studies
\cite{andersson2001,bakhtiariPRL2005,bakhtiariPoP2005,hirvijoki2015,decker2015}
have shown that the energy carried away by synchrotron and
bremsstrahlung radiation can be important in limiting the energy of
REs. This effect is especially pronounced during disruption mitigation
when large amounts of high-atomic-number gas (e.g. neon or argon) is injected, leading to
an increase in the effective charge of the plasma.

Kinetic modeling of electron runaway, accounting for the synchrotron
radiation reaction, is required to obtain an accurate description of
runaway electron distributions. Inclusion of the synchrotron radiation
reaction has been found to both limit the growth rate of the RE
population \cite{stahl2015} and lead to the formation of non-monotonic
features in the high energy tail of the distribution
\cite{hirvijoki2015,decker2015,pavel2015}. Similarly to synchrotron
radiation, bremsstrahlung radiation is expected to change the fast
electron distribution.

The reaction force on test particles has been extensively studied and
it was pointed out that bremsstrahlung and synchrotron radiation are
both effective in controlling the energy of runaway
electrons~\cite{bakhtiariPRL2005,bakhtiariPoP2005,solisetal2007}. 
A quantitative study of how bremsstrahlung affects the
runaway electron distribution function, however, is yet to be carried
out. In this work, we will recap the kinetic description of the
bremsstrahlung process in optically thin plasmas. This serves to
illustrate the physics involved, and also acts as a foundation for the
derivation of a simplified model operator, valid for typical runaway
scenarios and allowing a quantitative study of the effect of
bremsstrahlung on runaway electron dynamics. This, together with a
description of the linearized two-dimensional kinetic equation used in
the study, is contained in Section \ref{sec:bremsstrahlung}.

By extending the numerical tool \textsc{CODE} \cite{codepaper}, we
will in Section \ref{sec:Effect_on_dist} study the effect of
bremsstrahlung on the distribution of electrons in 2D
momentum-space. From the distribution we will deduce the impact on the
evolution of the fast-electron population, as well as how the synergy
between bremsstrahlung and synchrotron affect the dynamics. 
We show that, because of the sensitive interaction between
synchrotron force and pitch-angle scattering~\cite{hirvijoki2015},
a higher effective charge of the plasma does not 
tend to favor bremsstrahlung over synchrotron, unlike the results previously
reported~\cite{bakhtiariPRL2005,bakhtiariPoP2005,solisetal2007}.

In particular we derive a simple rule-of-thumb for when bremsstrahlung is
more important than synchrotron for the evolution of the runaway
population, and verify its validity numerically.  The plasma
parameters considered in this work primarily reflect those of electron
runaway in tokamak plasmas, but the validity of the model and the
conclusions (presented in Section~\ref{sec:conclusions}) are applicable to a broader range of plasma parameters.


\section{Bremsstrahlung radiation reaction}
\label{sec:bremsstrahlung}
The kinetic theory of inelastic processes is well understood, and is
described in detail in monographs on the subject such as
Ref.~\onlinecite{oxenius1986}.  Yet, in this section we shall provide
a brief but self-contained description of the kinetics of
bremsstrahlung reactions, presented in a form that is suitable for the
purposes of the present study.

Specifically, in this work, we consider fully ionized plasmas of
arbitrary ion composition near equilibrium, but with a population of
fast electrons present in the energy range $10-100\,$MeV, such as
those produced for example by runaway acceleration in tokamak
plasmas. In this energy range, reactions are accurately described only
by relativistic quantum theory. In this picture (in lowest-order
perturbation theory), bremsstrahlung is a binary interaction between
two charged particles, resulting in the emission of a photon.

The frequency of the bremsstrahlung radiation is assumed to far exceed
the plasma frequency, so that collective radiation effects can be
neglected, and the plasma is assumed to be optically thin. These
conditions impose no significant restriction; as the photon energies are
primarily of the same order as the incident electron energy, they
correspond to frequencies of the order $f \sim 10^{15} \,$Hz. The 
question of the optical depth will be briefly discussed in Sec.~\ref{sec:binary interactions}.

We start by describing the effect of binary interactions (also
referred to as collisions) on the rate of change of the distribution
function $f_a(t,\,\b{x},\,\b{p})$ of some particle species $a$ at time
$t$, position \b{x} and momentum \b{p}, defined such that
$n_a(t,\,\b{x}) = \int \rd \b{p} \, f_a(t,\,\b{x},\,\b{p})$ is the
number density of species $a$ at \b{x}, and
$N_a\{V\} = \int_V \rd \b{x} \, n_a(t,\,\b{x})$ is the total number of
particles $a$ in a volume $V$ at time $t$. From now on we suppress the
time- and space dependence of all functions as the collisions will be
assumed local in space-time, and we shall consider only spatially
homogeneous plasmas.

\subsection{Reaction rates}
\label{sec:binary interactions}
Bremsstrahlung reactions are inelastic Coulomb interactions in which a
photon is emitted with 4-momentum $k = (k/c,\,\b{k})$, with
$k=c|\b{k}|$ denoting the photon energy and $\b{k}$ the 3-momentum.
The process can be fully described by a differential cross-section
$\rd \sigma_{ab} = \rd
\sigma_{ab}(\b{p}_1,\,\b{p}_2,\,\b{k};\,\b{p},\,\b{p}'),$
determining the rate of interaction for collision events
$(\b{p},\,\b{p}') \mapsto (\b{p}_1,\,\b{p}_2,\,\b{k})$. Here \b{p} and
$\b{p}_1$ denote the momenta of particle $a$ before and after the
interaction, respectively, and similarly $\b{p}'$ and $\b{p}_2$ those
of particle $b$.
The case of elastic scattering corresponds to $\b{k}=0$. The
cross-section differential is over every degree of freedom of the
process with the initial momenta $\b{p}$ and $\b{p}'$ given,
constrained by conservation of momentum and energy. This can formally
be expressed as
\begin{align}
\rd \sigma_{ab} &= \delta^3(\b{p}+\b{p}' - \b{p}_1 - \b{p}_2 - \b{k}) \nonumber \\
&\times \delta(E+E'-E_1-E_2-k)A_{ab} \, \rd\b{p}_1\rd\b{p}_2\rd \b{k},
\label{eq:general cross section}
\end{align}
where the (often complicated) function $A_{ab}$ determines the
numerical value of the cross-section. The above expression can be
compared to Eq.~(2.1) in Ref.~\onlinecite{haug1975}, where
further details on manipulations of the bremsstrahlung cross-section are provided. After removing the delta functions by appropriate
integration over any four of the coordinates (one for each
conservation law), the remaining cross-section is a differential in five variables. One additional coordinate can be eliminated due to the azimuthal symmetry of the unpolarized
scattering process around the relative velocity of the incident
particles in the center-of-mass frame.

The change $(\rd n_a)_{c,ab}$ in the number density of species $a$ due
to collisions with particles of species
$b$ is given by~\cite{relativistic_boltzmann, akama1970}
\begin{align}
\hspace{-.4cm}\bigr(\rd n_a\bigr)_{c,ab} &= 
f_a(\b{p}_1) f_b(\b{p}_2) \bar{g}_{\text\o} \rd \bar{\sigma}_{ab} \rd\b{p}_1 \rd \b{p}_2 \rd t \nonumber \\
&\hspace{.4cm} -  f_a(\b{p})f_b(\b{p}')g_{\text\o} \rd \sigma_{ab} \rd \b{p} \rd \b{p}' \rd t.
\label{eq:collision rate}
\end{align}
Here
$g_{\text\o} = \sqrt{(\b{v}-\b{v}')^2 - (\b{v}\times \b{v}')^2/c^2}$
is the M\o{}ller relative speed, being the relativistic generalization
of the relative speed $v\sub{rel} = |\b{v}-\b{v}'|$; it correctly
reduces to $v\sub{rel}$ in the non-relativistic limit and in the
rest-frame of one particle its value is the speed of the other. In Eq.~\eqref{eq:collision rate}, a bar denotes a quantity for which we have let $\b{p}_1 \leftrightarrow \b{p}$ and $\b{p}_2 \leftrightarrow \b{p}'$, i.e.
$\rd\bar{\sigma}_{ab} = \rd
\sigma_{ab}(\b{p},\,\b{p}',\,\b{k};\,\b{p}_1,\,\b{p}_2)$,
and similarly for $\bar{g}_\text\o$.

The collision operator $C_{ab}$ describing the rate of change of the
distribution function at momentum \b{p} due to collisions is formally
introduced with the definition
\begin{align}
C_{ab}(\b{p}) \equiv \left(\frac{\partial f_a}{\partial t}\right)_{\!\!c,ab} = \int  \frac{(\rd n_a)_{c,ab}}{{\hspace{-4mm}\rd t \rd \b{p}}},
\label{eq:collision operator}
\end{align}
the integral to be taken over the variables indicated by
\Eq{eq:collision rate} after factoring out the momentum volume element
$\rd\b{p}$.  Equations (\ref{eq:collision rate}) and
(\ref{eq:collision operator}) fully determine the rate of change of
the distribution function at momentum \b{p} due to collisions, and we
can interpret the two terms in \Eq{eq:collision rate} as follows: The
first term, the gain term, expresses the rate at which particles $a$
are scattered \emph{into} \b{p} from $\b{p}_1$ due to all possible
interactions with particles $b$ of momentum $\b{p}_2$ in reactions
$(\b{p}_1,\,\b{p}_2) \mapsto (\b{p},\,\b{p}',\,\b{k})$. Likewise, the
second term, the loss term, expresses the rate at which particles $a$
are scattered \emph{away} from \b{p} due to all possible interactions
with particles $b$ of momentum $\b{p}'$, in interactions
$(\b{p},\,\b{p}') \mapsto (\b{p}_1,\,\b{p}_2,\,\b{k})$.

The differential cross-section can be split into separate contributions
from elastic and inelastic scattering:
$\rd \sigma = \rd \sigma^{\text{(elastic)}} +
\rd\sigma^{\text{(inelastic)}}$.
Elastic scattering between particles is described by
elastic Coulomb collisions, as the plasma has been assumed to be fully
ionized. The inelastic interactions are dominated by bremsstrahlung
reactions~\cite{heitler1954} which will be the focus of the remainder
of this section, and the form of the bremsstrahlung cross-section will
be given in Section~\ref{sec:cross-secs}. It will be illustrative to
briefly investigate the form of Eqs.~(\ref{eq:collision rate})
and (\ref{eq:collision operator}) for the elastic and inelastic cases separately, to
highlight the challenges in kinetic modeling of inelastic collisions.

\paragraph*{Elastic Coulomb collisions} --- For elastic processes ($\b{k}\!=\!0$),
\Eq{eq:general cross section} leaves only two variables when
integrated over the delta functions, allowing a differential
cross-section of the form
$\rd \sigma_{ab} = (\partial \sigma_{ab}/\partial \Omega) \rd \Omega$,
with $\Omega$ representing two scattering angles that parameterize any
elastic collision with given incident momenta.  We may here apply the
\emph{principle of detailed balance}, corresponding to the
time-reversibility of mechanical systems~\cite{johnjohn1969} or
quantum-mechanically to the symmetry of the transition matrix, valid
for all processes in the spin-averaged Born
approximation~\cite{heitler1954,weinberg}. It expresses the symmetry
of the reaction rates, and in the elastic case it takes the
form~\cite{relativistic_boltzmann}
\begin{align}
\bar{g}_{\text\o} \rd \bar{\sigma}_{ab} \rd\b{p}_1 \rd \b{p}_2 = g_{\text\o} \rd \sigma_{ab} \rd \b{p} \rd \b{p}'.
\end{align}
This allows \Eq{eq:collision operator} to be immediately written on
the symmetric form normally referred to as the Boltzmann collision
integral~\cite{landau kinetics, boltzmann1896},
\begin{align}
C_{ab}(\b{p}) = \int \rd \b{p}' \rd \Omega \, g_\text\o  \frac{\partial \sigma_{ab}}{\partial \Omega} \bigr[f_a(\b{p}_1)f_b(\b{p}_2)-f_a(\b{p})f_b(\b{p}')\bigr].
\end{align}
As elastic Coulomb collisions are often dominated by grazing
(small-angle, distant) collisions, the Boltzmann collision integral
above can be reduced to an operator of the Fokker-Planck
form~\cite{landau1936}. The Fokker-Planck operator we use to account
for elastic collisions in the plasma is described in
Section~\ref{sec:kinetic_eq_and_CODE}.

\paragraph*{Bremsstrahlung reactions} --- For bremsstrahlung, the
situation is different, and a collision operator on
the same symmetric form as for the elastic case can not be found.  By combining
Eqs.~(\ref{eq:collision rate}) and (\ref{eq:collision operator}) one
can always form the collision operator
\begin{align}
C_{ab}(\b{p}) &= \int \rd\b{p}_1\, f_a(\b{p}_1) \int \rd\b{p}_2 \, \bar{g}_\text\o f_b(\b{p}_2)\frac{\partial \bar\sigma}{\partial \b{p}} \nonumber \\
& \hspace{1.3cm} - f_a(\b{p})\int\rd\b{p}'\,g_\text\o f_b(\b{p}')\, \sigma, 
\label{eq:explicit collision operator}
\end{align}
where we introduced the two cross-sections
$\partial \bar\sigma/\partial \b{p}$ and $\sigma$, obtained by
integrating \Eq{eq:general cross section} over all remaining variables
on the right-hand side. For elastic collisions,
$\partial \bar\sigma/\partial \b{p}$ would retain one of the delta
functions, while for inelastic processes the collision operator must be given on the above form.

$C_{ab}$ is in general an integral operator containing six to eight integration variables,
depending on how easily $\partial \bar\sigma/\partial \b{p}$ can be
obtained. From a numerical point of view, evaluation of the bremsstrahlung cross-sections are often demanding due to large cancelations between terms in the limits of small emission angle  or high or low photon energy. In addition, both
$\partial\bar\sigma/\partial \b{p}$ and $\sigma$ are generally
logarithmically divergent expressions in the low-photon energy
contributions (physically corresponding to a large number of
low-energy photons being emitted, carrying a finite total
energy). While the last issue can be resolved by cutting the integral
off at photon wavelengths comparable to the dimension of the system,
the evaluation of the cross-section is often computationally expensive.

Due to the infeasibility of integrating a detailed description of the
bremsstrahlung process into current kinetic-equation solvers (because
of its computational intensity), we will in the next section introduce
a simplified model of the bremsstrahlung collision operator in
\Eq{eq:collision operator}, retaining some of its essential physical
features while neglecting others. This will allow efficient solution of
a kinetic equation suitable for runaway electrons, letting us quantify
the importance of bremsstrahlung in magnetized plasmas where it, as an
energy loss mechanism, competes primarily with synchrotron radiation.

Finally, to clarify the difference between the elastic and inelastic
collision operators, the reason that the principle of detailed balance
does not apply in the inelastic case is that it relates the bremsstrahlung
cross-section to the reverse process where a photon is absorbed
(inverse bremsstrahlung). However, as we assume the plasma to be
optically thin, the emitted photons will promptly leave the plasma
before interacting again with the energetic electrons. The validity of
this assumption is readily verified: the multi-MeV bremsstrahlung photons (having frequencies $\omega \gg \omega_p$ much
larger than the plasma frequency) primarily interact with the fast
electrons through Compton scattering, for which the photon mean free
path is of the order
$\lambda \gtrsim 1/ (n_{\text{RE}} r_0^2) \approx
(10^{29}\,\text{m}^{-2})/n_{\text{RE}}$,
where $r_0 = e^2/(4\pi\varepsilon_0 m_e c^2) $ is the classical
electron radius. As we are interested in runaway electron beams with
typical densities $n_{RE} \ll 10^{20}$ m$^{-3}$ and typically
populating space with linear dimension of order $L = 1$-$10\,$m, it is
clear that emitted photons will not significantly interact again with
the runaway electrons (the ratio of beam dimension and photon mean
free path $\lambda$ being $L/\lambda \ll 10^{-9}$). We may therefore
neglect the presence of the photon distribution for the present study
of runaway electron dynamics.

\subsection{Single-particle model operator for bremsstrahlung}

Due to the complexities of fully modeling bremsstrahlung interactions in the
kinetic equation, we will seek a simpler form. \emph{Ad-hoc}, we
will assume a certain analytic form for the collision operator, and to uniquely specify its value we will demand that certain conditions -- consistent with properties of the full collision operator in Eqs.~(\ref{eq:collision rate}) and (\ref{eq:collision operator}) -- are satisfied.  As we are interested in investigating the competition
between accelerating electric field and bremsstrahlung as a
slowing-down mechanism, it will be useful to look for a bremsstrahlung
operator in the form of an effective single-particle force. Such a
force would take the form of an operator
\begin{align}
C^{(m)} = -\frac{\partial}{\partial \b{p}}\cdot\Big(\b{F}\sub{B}(\b{p}) f_e(\b{p})\Big),
\label{eq:model operator}
\end{align}
where $\b{F}_B$ can be identified as the reaction force associated
with the bremsstrahlung interactions. A condition that uniquely
defines $\b{F}_B$ can be found by requiring that the model operator
$C^{(m)}$ produces the same total radiated energy as the full
collision operator for an arbitrary distribution function. The energy
rate of change is defined as (using
$\sqrt{m_e^2c^4 + p^2 c^2} = \gamma m_e c^2$ for the particle energy)
\begin{align}
W_{c,a} &\equiv \int \rd \b{p} \, m_e c^2 \gamma \left(\frac{\partial f_a}{\partial t}\right)_c \nonumber \\
& = m_e c^2 \sum_b \int \, \gamma \, (\rd n_a)_{c,ab} / \rd t.
\end{align}

Due to the symmetry in performing the integration over all variables, we may let
$(\b{p}_1,\,\b{p}_2) \leftrightarrow (\b{p},\,\b{p}')$ in the first term of \Eq{eq:collision rate}, yielding
\begin{align}
W_{c,a} = -m_e c^2 \sum_b \int \rd \b{p}\rd\b{p}' \rd\sigma \, g_\text\o (\gamma - \gamma_1) f_a(\b{p})f_b(\b{p}').
\end{align}
The large mass difference between ions and electrons allows approximations to be made when calculating the energy radiated by the electron population, making it advantageous to treat the electron-ion and electron-electron interactions separately. In the electron-ion
contribution, the characteristic speed of an ion in a population near
equilibrium is much smaller than the electron velocities,
$v_{Ti} = \sqrt{2T_i/m_i} \ll \sqrt{2T_e/m_e} = v_{Te}$, assuming
$T_e$ and $T_i$ to be of the same order. Therefore, as the
differential cross-section is a function of the center-of-mass energy
and hence varies over scales of order the incident electron momentum,
we may approximate the ion population as delta-distributed,
$f_i(\b{p}') = n_i\delta(\b{p}')$. Conservation of energy allows us to
relate $m_e c^2(\gamma-\gamma_1) = k$ when neglecting ion recoil
effects, which are formally of order $\gamma m_e/m_i$. Consequently, we obtain a
simpler relation for the energy loss rate:
\begin{align}
W_{c,e\text{-}i} = - \sum_i n_i \int \rd \b{p} \,  v f_e(\b{p}) \int \rd \sigma_{e\text{-}i} \, k.
\label{eq:e-i energy rate}
\end{align}
For electron-electron interactions, the expression in \Eq{eq:collision
  rate} is symmetric with respect to permutations of
$(\b{p}_1,\,\b{p}_2)$ and $(\b{p},\,\b{p}')$ when integrated, as we
now consider identical particles. Therefore, we may replace $\gamma$
by $(\gamma+\gamma')/2$ and $\gamma_1$ by $(\gamma_1+\gamma_2)/2$, yielding the energy rate
\begin{align}
W_{c,e\text{-}e} = -\frac{1}{2}\int \rd \b{p}\rd\b{p}'\,g_\text\o f_e(\b{p})f_e(\b{p}') \int \rd \sigma_{e\text{-}e} \, k,
\label{eq:electron energy rate}
\end{align}
where now $k = \gamma + \gamma' - \gamma_1 - \gamma_2$.  This can be
simplified to a form similar to the electron-ion case under the
assumption that the fast-electron population can be treated as a
perturbation to a sufficiently cold equilibrium. In that case, we may
write $f_e = f_{e0} + f_{e1}$, where $f_{e0}$ is the equilibrium
solution and the perturbation $f_{e1}$ includes the fast-electron
population. In \Eq{eq:electron energy rate} we then find the term
$f_e(\b{p})f_e(\b{p}') = f_{e0}(\b{p})f_{e0}(\b{p}') +
f_{e0}(\b{p})f_{e1}(\b{p}') + f_{e1}(\b{p})f_{e0}(\b{p}') +
f_{e1}(\b{p})f_{e1}(\b{p})$.
The last term, quadratic in the perturbation, can be neglected as
$f_{e1}$ is assumed small.  In addition, the $e$-$e$ bremsstrahlung
cross-section vanishes in the limit where both incident electron
velocities are much smaller than the speed of light, $v,\,v' \ll c$,
as in that limit bremsstrahlung reduces to dipole radiation, with a
two-electron system having zero dipole moment. Therefore, as long as
the electron temperature satisfies $T_e \ll m_e c^2$, the first term,
quadratic in the equilibrium distribution, is
negligible~\cite{haug1975}. The remaining two terms can be combined
due to the symmetry between $\b{p}$ and $\b{p}'$ inside the integral,
leaving the expression
\begin{align}
W_{c,e\text{-}e} = -n_e\int \rd \b{p}\, v f_e(\b{p}) \int \rd \sigma_{e\text{-}e} \, k.
\end{align}
Here, it has further been assumed that
$f_{e0}(\b{p}) = n_e \delta(\b{p})$, since the energies of the fast
electrons far exceed the thermal energy.

An elementary calculation shows that the energy moment of the model
operator given in \Eq{eq:model operator} is
\begin{align}
W^{(m)} &= \int \rd\b{p} \, \gamma m_e c^2 C^{(m)} = \int \rd\b{p} \,  m_e c^2 \b{F}\sub{B}\cdot\frac{ \partial \gamma}{\partial \b{p}} f_e \notag \\ &= \int \rd\b{p} \,  (\b{v}\cdot\b{F}_B) f_e ,
\end{align}
giving the same answer as $\sum_b W_{c,e\text{-}b}$ for any
distribution function $f_e$, provided that we choose the bremsstrahlung
reaction force as
\begin{align}
\b{F}_B = - \hat{p}\sum_b n_b \int \rd \sigma_{e\text{-}b} \, k.
\label{eq:bremsstrahlung force}
\end{align}
While a component orthogonal to the momentum unit vector
$\hat{p}=\b{p}/p$ would be consistent with the condition
$W^{(m)} = W_{c,e}$, it can be ruled out on a physical basis since the
force arises from the interaction with spherically symmetric
background particle populations (delta-distributed in our linearized
high-energy limit treatment), hence lacking any preferential direction
except for that of the particle motion. By the same argument the force
must also be independent of pitch-angle, and therefore represents
isotropic friction.

A few remarks are in order regarding the approximation used above. The
form in \Eq{eq:bremsstrahlung force} for the bremsstrahlung force is
commonly employed as a simple model for electron slowing-down by
bremsstrahlung~\cite{bakhtiariPRL2005, bakhtiariPoP2005,haug2004},
corresponding to the so-called radiation cross section often found in
the literature \cite{heitler1954,jackson}.

While this form for the reaction force is both well-established and
elementary, the derivation provided here illustrates the conditions
for its validity, and also highlights the physics that is lost: in
treating the bremsstrahlung collision process in the manner described
above, two effects have been neglected. First, it is clear that the
force operator in \Eq{eq:model operator} describes only a continuous
slowing-down of the electrons. Since the photons emitted in
bremsstrahlung interactions often have energies of the same order as
the incident electron, the slowing-down would in reality occur in
leaps rather than continuously. In addition, as the photons are
emitted in random directions and momentum is transferred to the
target, the electrons are also deflected in the process (corresponding
to a change in the pitch-angle), even though for an ensemble of
electrons the angular deviation averages to zero. This would, akin to
elastic Coulomb scattering, diffuse the electron distribution in
momentum space, an effect which is not captured in the above
treatment.

Finally, we remark that no energy is transfered from the electrons to
the ions (which are assumed to be infinitely heavy), making \Eq{eq:e-i energy
  rate} an accurate measure of the energy lost from the fast-electron
population due to electron-ion bremsstrahlung radiation. However, in
electron-electron interactions, energy will be transferred from the
energetic electron not only to the photon, but also to the slow target
particle. Therefore \Eq{eq:electron energy rate} is a lower bound for
the energy loss of the fast-electron population. This can be compared
to elastic scattering: since in that case $k\equiv 0$, the integral in
\Eq{eq:electron energy rate} would vanish identically. However,
electron slowing-down still occurs due to the energy transfer to the
slow target electron, which would not be captured by this treatment.

Also note that, due to conservation of energy and since no energy
is transfered to the ion species, $W_{c,e}$ denotes the total
power emitted from the plasma as (bremsstrahlung) X-rays. By
construction this is true also for $W^{(m)}$, limited only by the
validity of the linearization procedure outlined above.


\subsection{Cross-sections for bremsstrahlung emission}
\label{sec:cross-secs}
In this section we give explicit expressions for the bremsstrahlung
reaction force. Due to the fundamental limitations of the model
operator derived in the previous section, we will seek no greater
accuracy in the cross-sections than that provided by the Born
approximation in relativistic quantum theory. As we only consider
fully ionized plasmas, we do not need to account
for atomic form factors that would modify the formulas in the presence
of bound electrons.

\paragraph*{Electron-ion collisions}--- The cross-section for
electron-ion bremsstrahlung interactions (with ions of charge $Z$) is
given by the Bethe-Heitler formula \cite{betheheitler1934}
\begin{align}
\rd^4\sigma &= \frac{\alpha Z^2 r_0^2}{2\pi}
\frac{p \sin\theta \sin\theta_0 \rd k \rd \theta \rd \theta_0 \rd \varphi}{p_0 k q^4} \notag \\
&\ \times \Biggr\{  \frac{p^2 \sin^2\theta}{(E-p\cos\theta)^2}(4E_0^2 - q^2) \notag \\
&\qquad + \frac{p_0^2 \sin^2 \theta_0}{(E_0-p_0\cos\theta_0)^2}(4E^2-q^2) \nonumber \\
&\qquad - \frac{2p p_0\sin\theta\sin\theta_0 \cos\varphi\,(4E_0E-q^2+2k^2)}{(E-p\cos\theta)(E_0-p_0\cos\theta_0)} \nonumber \\
&\qquad + 2k^2 \frac{p^2\sin^2\theta + p_0^2\sin^2\theta_0}{(E-p\cos\theta)(E_0-p_0\cos\theta_0)} \Biggr\},
\end{align}
where $q = |\b{p}_0-\b{k}-\b{p}|$ denotes the magnitude of the
momentum transferred to the ion,
$r_0 = e^2/(4\pi\varepsilon_0 m_e c^2) \approx 2.8\cdot 10^{-15}\,$m
is the classical electron radius,
$\alpha = r_0 m_e c/\hbar \approx 1/137$ is the fine-structure
constant,
$e$ is the magnitude of the elementary charge, $\varepsilon_0$ is the
vacuum permittivity, and all momenta and energies are normalized to
$m_e c$ and $m_e c^2$, respectively. Subscript 0 denotes the incident
electron, no subscript the outgoing electron, and $k$ is the photon
energy. We define $\theta_0$ ($\theta$) as the angle between incident
(outgoing) electron and the emitted photon, while $\varphi$ is the
azimuthal angle (around the axis $\b{k}$) between incident and
outgoing electron. The ion is assumed to be an infinitely massive
scattering center, meaning that order $\gamma m_e/m_i$ recoil effects
have been neglected.

This formula can be integrated analytically~\cite{racah1934} to give 
the electron-ion bremsstrahlung reaction force (the sum over ion species 
is trivial since the ion parameters only appear in the charge factor $Z_i^2$ multiplying the cross-section),
 \begin{align}
F_{\text{B},e\text{-}i}(p) &= - \sum_i n_i \int \rd \sigma_{e\text{-}i} \, k \nonumber \\ 
&= - \alpha Z\sub{eff} n_e r_0^2 m_e c^2 (\gamma - 1) \nonumber \\
&\ \times \Biggr( 
 \frac{4}{3}\frac{3 \gamma^2 + 1 }{\gamma p} \ln(\gamma+p) - \frac{(8\gamma + 6p)}{3\gamma p^2}\Big(\ln(\gamma+p)\Big)^2 \notag \\
&\qquad - \frac{4}{3} + \frac{2}{\gamma p} \int_0^{2p(\gamma+p)} \hspace{-2mm} \rd x\, \frac{\ln(1+x)}{x} \Biggr),
\end{align}
where $Z\sub{eff} = \sum_i n_i Z_i^2/n_e$ is the effective charge of the plasma.
In the high energy limit where $p^2 \approx \gamma^2 \gg 1$, this expression takes the asymptotic form

\begin{align}
F_{\text{B},e\text{-}i}(p) &= - \frac{\alpha}{\pi}\frac{eE_c Z\sub{eff}}{\ln\Lambda}(\gamma-1) \left(\ln2\gamma - \frac{1}{3}\right), 
\label{eq:F_Bii}
\end{align}
where $E_c = n_e e^3\ln{\Lambda} /4\pi\varepsilon_0^2 m_e c^2 = 4\pi \ln\Lambda n_e r_0^2 m_e c^2 / e$ is the critical electric field for runaway electron generation\cite{connorhastie1975}, defined as the minimum electric field
above which runaway electrons can be generated in the absence of loss mechanisms other
than collisional friction, and $\ln\Lambda$ is the Coulomb logarithm.


\paragraph*{Electron-electron collisions}--- For $e$-$e$ collisions,
the full expression for the bremsstrahlung cross-section valid at all
energies is unwieldy~\cite{elementary}.  We will instead
use the high-energy limit which, when integrated, yields a
bremsstrahlung force of the same form as \Eq{eq:F_Bii} but with
$Z\sub{eff}=1$.

\begin{figure}[b]
\begin{center}
\includegraphics[width=.5\textwidth]{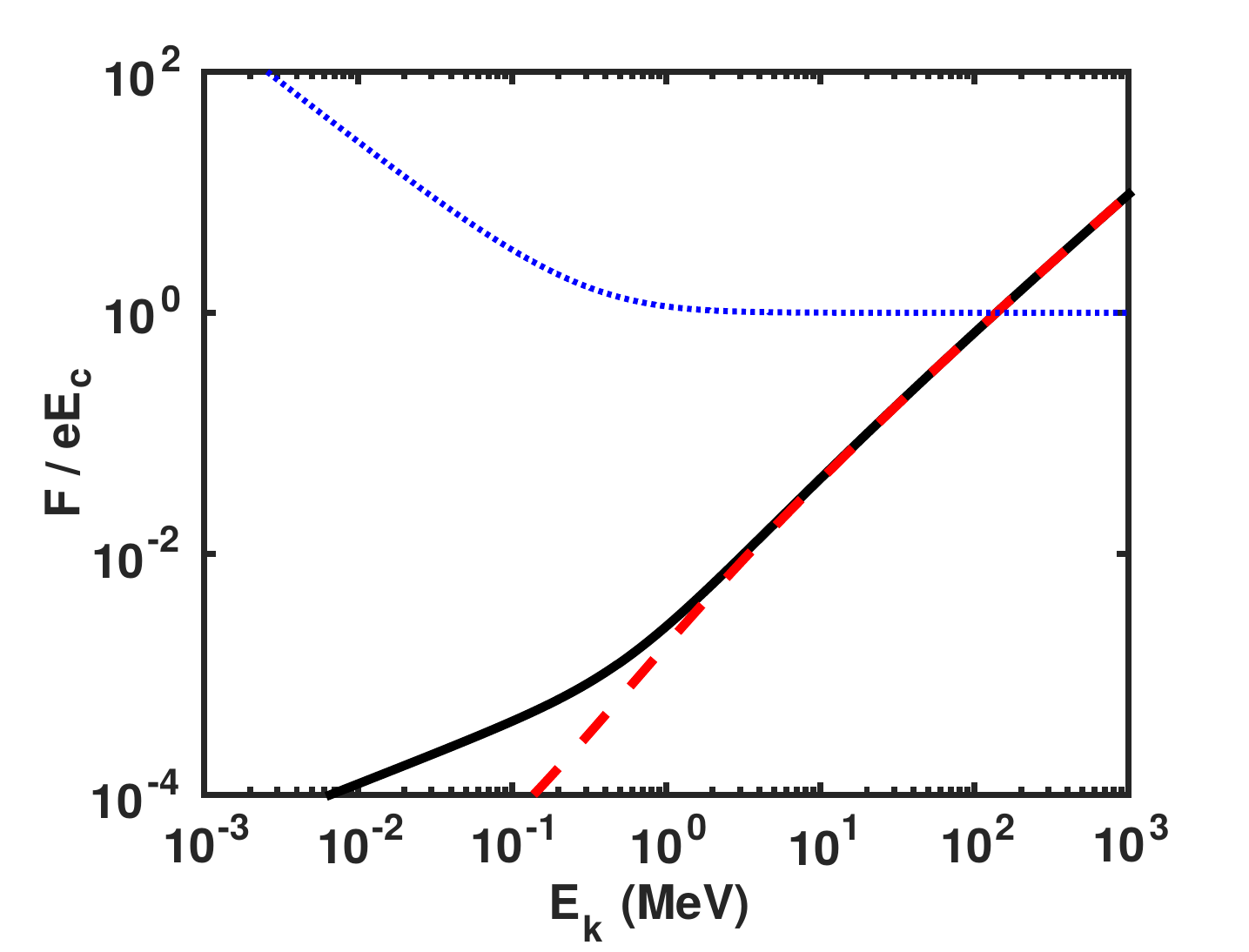}
\caption{\label{fig:force benchmark} Comparison of collisional energy losses in a plasma with constant 
$\ln\Lambda = 15$ and $Z\sub{eff}=3$. The lines represent numerical integration of the full 
Born-approximation bremsstrahlung radiation cross-section~\cite{haug2004} (black, solid), the 
bremsstrahlung force used in this work (red, dashed), 
and the dynamical friction force associated with elastic Coulomb collisions (blue, dotted).}
\end{center}
\end{figure}

We then obtain for the full bremsstrahlung reaction, accounting for electron-ion and electron-electron interactions, 
\begin{align}
F\sub{B}(p) =-\frac{\alpha(1+Z\sub{eff})eE_c}{ \pi \ln\Lambda } (\gamma-1)(\ln 2\gamma - 1/3).
\label{eq:F_a}
\end{align}
This is the formula commonly used in the literature for the electron bremsstrahlung stopping power, for example given in Ref.~\onlinecite{betheheitler1934}, and used in Refs.~\onlinecite{bakhtiariPRL2005, bakhtiariPoP2005, solisetal2007} to study electron slowing-down in plasmas. We can compare this expression for the force to a more accurate formula derived in Ref.~\onlinecite{haug2004}, which was developed as a fit to numerical evaluations of the full bremsstrahlung cross-section in the Born approximation. These expressions are shown in \Fig{fig:force benchmark} for the case $Z\sub{eff}=3$ and $\ln\Lambda = 15$, together with the friction force due to elastic Coulomb collisions. 

It is seen that the agreement between our high-energy limit formula and the more accurate expression is excellent in the entire region where bremsstrahlung losses are significant compared to collisional friction. Due to the rapid fall-off of the bremsstrahlung force at low energy, the minimum friction is barely modified by the inclusion of bremsstrahlung losses, hence the critical electric field for runaway generation should largely be unaffected by its addition (certainly negligible within the order $1/\ln\Lambda$ accuracy of the Fokker-Planck equation).


\subsection{Kinetic equation and implementation in \textsc{CODE}}
\label{sec:kinetic_eq_and_CODE}
We wish to describe the time evolution of fast-electron
populations under the influence of elastic and inelastic collisions,
quasi-static electric fields and radiation reaction forces. We shall
consider only plasmas where the Coulomb logarithm is large
($\ln\Lambda \gtrsim 10$) such that a Fokker-Planck treatment of the
elastic Coulomb collisions is appropriate.

For the runaway problem we use a relativistic 0D+2V kinetic (Fokker-Planck) equation for the electron distribution,
\begin{align}
\frac{\partial f_e}{\partial t} + \left\langle\frac{\partial}{\partial \b{p}}\cdot \bigr[ (\b{F}\sub{L} + \b{F}\sub{S} + \b{F}\sub{B}) f_e \bigr]\right\rangle
= C_{ei} + C_{ee}.
\label{eq:kinetic equation}
\end{align}
Here $\b{p} = \gamma m_e \b{v}$ is the relativistic electron 3-momentum, 
$\b{F}\sub{L}$ is the Lorentz force, while $\b{F}\sub{S}$ and $\b{F}\sub{B}$ denote the radiation reaction forces associated with 
synchrotron and bremsstrahlung radiation, respectively. The collision operators $C_{ei}$ and $C_{ee}$ describes the effect of elastic Coulomb 
collisions with ions and electrons. As we consider magnetized plasmas, the equation has been 
averaged over the gyro-motion (represented by the brackets), with
\begin{align}
\left\langle\frac{\partial}{\partial \b{p}}\cdot \bigr(\b{F}\sub{L} f_e\bigr) \right\rangle &= -e E_\parallel \left(\frac{\partial}{\partial p} + \frac{1-\xi^2}{p}\frac{\partial}{\partial \xi}\right)f_e, \\
 \left\langle\frac{\partial}{\partial \b{p}}\cdot \bigr(\b{F}\sub{S} f_e\bigr) \right\rangle &= -\frac{1}{\tau\sub{S}} \biggr[ \frac{1}{p^2}\frac{\partial}{\partial p}\left( \gamma  p^3 [1-\xi^2] f_e \right) \nonumber \\
 & \hspace{12mm}+ \frac{1}{p}\frac{\partial}{\partial \xi}\left(  \frac{p\xi}{\gamma}[1-\xi^2] f_e \right) \biggr], \\
\left\langle\frac{\partial}{\partial \b{p}}\cdot \bigr(\b{F}\sub{B} f_e\bigr) \right\rangle &= -\frac{m_e c}{\tau\sub{B}} \frac{1}{p^2}\frac{\partial}{\partial p}\left[ p^2(\gamma-1)(\ln 2\gamma - 1/3)f_e \right],
\end{align}
where $p = |\b{p}|$, \b{E} is the electric field and the subscript $\parallel$ denotes the Cartesian component parallel to the magnetic field, $\xi = p_\parallel / p$ is the pitch-angle cosine, and the characteristic radiation reaction time scales $\tau\sub{S}$ and $\tau\sub{B}$ for synchrotron and bremsstrahlung respectively are
\begin{align}
\tau\sub{S} &= \frac{6\pi\varepsilon_0 (m_e c)^3}{e^4 B^2}, \\
\tau\sub{B} &= \frac{\pi \ln\Lambda  \, m_e c }{\alpha(1+Z\sub{eff})eE_c} = \frac{\pi \ln\Lambda}{\alpha(1+Z\sub{eff})}\tau_c,
\end{align}
where $B$ is the magnetic field strength and $\tau_c = m_e c/ (e E_c)$
is the characteristic time scale of elastic collisions for relativistic electrons
of energy much greater than the thermal energy. A detailed derivation
of this particular form of the kinetic equation can be found, e.g., in
Ref.~\onlinecite{hirvijoki2015}. Note that the same kinetic equation
is valid also for cylindrically symmetric unmagnetized  plasmas ($B=0$) , with
$\parallel$ then denoting the direction of the electric
field. Equation~(\ref{eq:kinetic equation}) has been implemented in
the numerical initial-value Fokker-Planck solver
\textsc{CODE}~\cite{codepaper}, which is used for all numerical
results presented in this work.

As we are interested in the momentum space dynamics of the runaway
electrons, only the test-particle part of the linearized
electron-electron collision operator $C_{ee}$ is retained. The ions
are assumed to be stationary and infinitely massive, and $C_{ei}$
therefore only describes pitch-angle scattering, while $C_{ee}$ is an
asymptotic matching of the non-relativistic test-particle operator and
the high-energy limit of the relativistic test-particle operator,
yielding good agreement with the exact operator across the full energy
range.

In the following, we will use a parameter $\sigma$ to
characterize the relative strength of the synchrotron reaction force
compared to collisions. It is defined as the ratio of the
characteristic time-scales, and is given by
\begin{align}
\sigma = \frac{\tau_c}{\tau\sub{S}}=\frac{2\varepsilon_0}{3m_e}\frac{B^2}{n_e \ln\Lambda}.
\label{eq:sigma}
\end{align}


\section{General effect of radiation reaction on the electron distribution function}
\label{sec:Effect_on_dist}

It was recently
shown that the inclusion of the synchrotron radiation reaction force
in the kinetic equation could induce the formation of non-monotonic
features (a ``bump") in the high energy tail of the electron
distribution\cite{hirvijoki2015, decker2015}. The formation of such
bumps is a consequence of the interplay between an accelerating force
(the electric field, $E\!>\!E_c$) and the presence of an effective
friction which increases with particle energy and
pitch-angle\cite{decker2015}. The presence of the bump effectively
limits the maximum energy obtained by the runaways. Since our 
bremsstrahlung radiation reaction force is isotropic, it
does not exhibit the same pitch-angle dependence as the synchrotron
force. However, it will act to reduce the particle momentum, thereby
also limiting the maximum attainable particle energy. Therefore, 
inclusion of bremsstrahlung radiation reaction is also expected to 
result in the formation of a bump.

\begin{figure}
	\begin{center}
		\includegraphics[width=1\columnwidth, trim={0 0.67cm 0.3cm 1.05cm},clip]{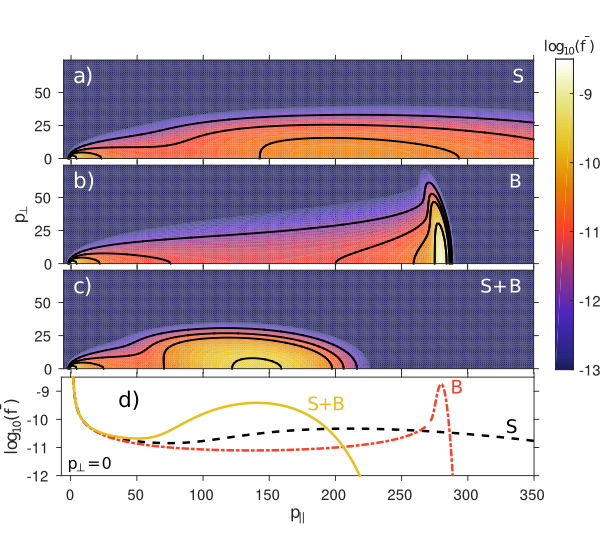}
		\caption{\label{fig:dists_S_B_SB} Contour plots of
                  electron distributions in 2D momentum space,
                  exhibiting bumps induced by radiation
                  reaction. Included reaction forces were a)
                  synchrotron, b) bremsstrahlung and c) both
                  synchrotron and bremsstrahlung. Cuts at
                  $p_\perp\!=\!0$ are shown in d). The distributions
                  were obtained using \textsc{CODE} for the parameters
                  $T_e=10$~keV, $n_e=2.3\cdot 10^{21}$~m$^{-3}$,
                  $B=4$~T, $\zeff=3$ and $E/E_c=2$, after
                  $t=5\cdot 10^5\tau_{ee}$. The plotted quantity is
                  $\log_{10}(\bar{f})$, where $\bar{f}$ is the
                  distribution normalized so that
                  $\bar{f}(p\!=\!0)=1$.}
	\end{center}
\end{figure}

Figure \ref{fig:dists_S_B_SB} shows examples of the electron
distribution function, calculated using \textsc{CODE}, with
synchrotron (S), bremsstrahlung (B), and both synchrotron and
bremsstrahlung (S+B) radiation reaction effects
included. Non-monotonic features are clearly formed in all three
cases, but their characteristics are significantly different. In the
case of synchrotron radiation only, the bump is not very pronounced
and does not significantly restrict the runaway energy (for the
parameters used), leading to a drawn-out tail. In contrast, the
bremsstrahlung-only bump is a sharp feature, accompanied by
strong gradients in $p$ and significant spreading in pitch. The
combined effect of synchrotron and bremsstrahlung radiation reaction
leads to a bump that has a large extension in $p_{||}$, but which is
clearly defined and efficiently limits the maximum particle energy. It
thus exhibits features from both the pure-synchrotron and
pure-bremsstrahlung bumps, but is accompanied by a strong reduction in
the particle energy at the bump peak (what we will refer to as the
\emph{bump location}), compared to the other two cases.

The dynamics in the presence of both synchrotron and bremsstrahlung
forces can be understood by the following argument: the synchrotron
reaction force, proportional to $p_\perp^2$ at large energies, acts
mainly to limit the perpendicular momentum of the fast electrons,
while allowing them to accelerate in the parallel direction. However,
once the electrons reach an energy for which $F\sub{B} + |eE_\parallel| <0$
the spherically symmetric bremsstrahlung reaction force will stop them
from being further accelerated in the parallel direction.  As the
electron distribution builds up in this region of momentum space, the
increasing perpendicular gradient will cause a diffusive perpendicular
momentum flux of electrons due to pitch-angle scattering, making the
synchrotron force increasingly significant. This suppresses further
expansion in the perpendicular direction.

The distributions shown in \Fig{fig:dists_S_B_SB} are no longer
changing -- the system has evolved long enough for the distribution in
Figs.~\ref{fig:dists_S_B_SB}b and c to reach steady-state. In
\Fig{fig:dists_S_B_SB}a, the distribution is in a quasi-steady
state. In this case, the numerical grid is not large enough to contain
the complete distribution, and as a consequence there is a constant
outflow through the boundary of the simulation domain.

\section{Relative importance of bremsstrahlung and synchrotron radiation reaction}
\label{Sec:Relative_importance}

\subsection{Analytical estimate}
\label{sec:Estimate_BS_vs_B}
As shown in the previous section, both synchrotron and brehmsstrahlung
radiation reaction can have a strong effect on the distribution function.
In many scenarios, however, one of them will be dominant.
It is thus informative to find an approximate condition to determine
the relative importance of these mechanisms. This may be accomplished
by comparing the location of the local maximum produced in the runaway
tail for the two radiation-reaction forces separately (while
neglecting the effect of the other).  For energies greater than this
maximum, the distribution exhibits a rapid decay with energy; hence
the location of the maximum can be regarded as an approximate upper
limit for the electron energy (for the case of pure synchrotron
losses, the parallel width of the bump can be significant as illustrated in
\Fig{fig:dists_S_B_SB}, but the decay is asymptotically exponential
with energy~\cite{hirvijoki2015} and the probability of finding
particles with energies much larger than the location of the maximum is low).

In Ref.~\onlinecite{hirvijoki2015}, the particle momentum at the
location of the maximum of the bump formed in the runaway tail due to
the synchrotron reaction force was investigated.  In the high energy
limit, a lower bound in $p$,
\begin{align}
p_{0\text{S}} > \frac{1+\sigma}{\sigma}\frac{E/E_c-1}{1+Z\sub{eff}}, \label{eq:p0S}
\end{align}
was derived (where $\sigma$ is given by Eq.~\ref{eq:sigma}). While the
synchrotron force does not depend on $Z\sub{eff}$ directly (since it
is not a collisional effect), the synchrotron power radiated by an
electron depends strongly on its pitch angle.  The effective
synchrotron force on a distribution of electrons is therefore directly
proportional to the pitch-angle scattering term in the kinetic
equation (as demonstrated in Ref.~\onlinecite{hirvijoki2015} and
exhibited by Eq.~\ref{eq:p0S}), which does scale with the effective
charge.

\begin{figure}
	\begin{center}
		\includegraphics[width=1\columnwidth, trim={0.2cm 0 0.8cm 0.3cm},clip]{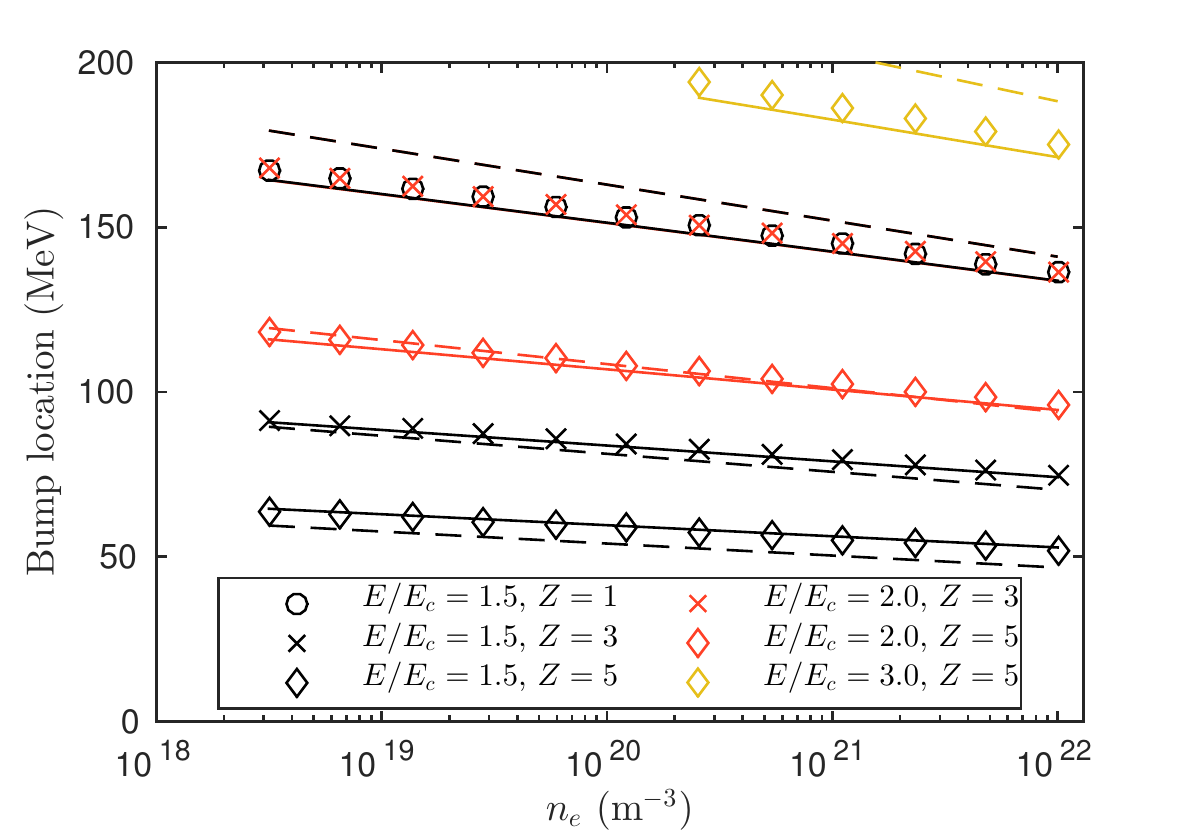}\\
		\caption{\label{fig:brems_bump_location} Location of
                  the bump in the distribution tail, induced by
                  bremsstrahlung radiation reaction, for various
                  combinations of density, electric field strength and
                  effective charge. Symbols represent distributions
                  obtained using \textsc{CODE} after
                  \mbox{$t=5\cdot 10^5\tau_{ee}$}, with $T_e=10$ keV
                  and neglecting synchrotron radiation reaction. The
                  dashed and solid lines are the corresponding
                  estimates for the bump location given by
                  \Eq{eq:brems_bump_location}, using $\gamma_0$
                  corresponding to 100 MeV and taken from the location
                  of the \textsc{CODE} bump, respectively. The
                  agreement when using the more accurate value for
                  $\gamma_0$ is excellent, but also the more rough
                  value gives a good estimate for the actual bump
                  location.}
	\end{center}
\end{figure}

We can find a similar estimate for the maximum electron energy in the
presence of bremsstrahlung (neglecting synchrotron effects) by
considering force balance on a test particle. There are primarily three 
forces acting on an electron in the high energy limit: the accelerating force of the parallel
electric field, collisional friction, and the bremsstrahlung reaction
force. The maximum electron energy can be expected to approximately
coincide with the momentum for which the forces balance, above which
friction will be greater than the accelerating electric field. This
yields the condition
\begin{align}
eE - eE_c - \frac{1+Z\sub{eff}}{\pi\ln\Lambda}\alpha eE_c\gamma(\ln 2\gamma- 1/3) = 0,
\end{align}
where we have used the high energy limit of both the bremsstrahlung
reaction and collisional friction forces. We may find a simple
expression by recognizing that in the last term, the logarithm varies
slowly with energy and may be replaced by a characteristic value
$\ln 2\gamma_0$, where $\gamma_0$ represents a typical energy scale
for the local maximum of the runaway tail. Solving for
$p_{0\text{B}}\approx \gamma$ gives
\begin{align}
p_{0\text{B}} = \frac{\pi \ln\Lambda}{\alpha(\ln 2\gamma_0-1/3)}\, \frac{E/E_c-1}{1+Z\sub{eff}}. \label{eq:brems_bump_location}
\end{align}
This estimate for the location of the bump shows good agreement with
distributions obtained using \textsc{CODE}, as demonstrated in
\Fig{fig:brems_bump_location}.

It can be expected that bremsstrahlung will have a significant effect
on the distribution (as compared to synchrotron radiation) when
$p_{0B} \lesssim p_{0S}$, giving the condition
\begin{align}
  \frac{\pi\ln\Lambda}{\alpha(\ln 2\gamma_0-1/3)} \lesssim \frac{1+\sigma}{\sigma}.
\end{align}
The left-hand side is typically of order $10^3$, thus corresponding to
small values of $\sigma$. Therefore it is justified to neglect the
term $\sigma$ appearing in the numerator, and by inserting its
definition from Eq.~(\ref{eq:sigma}) we obtain a condition relating
the magnetic field strength and electron density,
\begin{align}
n_e \gtrsim \frac{2\pi}{3\alpha(\ln 2\gamma_0-1/3)}\frac{\varepsilon_0}{m_e} B^2.
\end{align}
With the electron density expressed in units of $10^{20}\,$m$^{-3}$
and magnetic field expressed in Tesla, we find that bremsstrahlung
effects become significant in limiting the maximum energy of runaway
electrons for densities above
\begin{align}
n_{e,\text{lim}}[10^{20}\,\text{m}^{-3}] \approx \frac{28}{\ln 2\gamma_0-1/3} B[T]^2. \label{eq:n20lim}
\end{align} 
For a typical location of the maximum electron energy at around
$E_k = 100\,$MeV, we obtain $n_{e,\text{lim}}[10^{20}\,\text{m}^{-3}] \simeq 5 B[T]^2$.
In Ref.~\onlinecite{solisetal2007} a similar condition was derived,
showing a similar order of magnitude on the density $n_{e,\text{lim}}$, but
incorrectly suggesting that bremsstrahlung will be more important for 
larger $Z\sub{eff}$. The reason for this is that the test-particle picture 
employed in Ref.~\onlinecite{solisetal2007} does not accurately describe the pitch-angle scattering, hence
the resulting synchrotron emission is underestimated.

In present day tokamak experiments, $n_e \lesssim 10^{20}\,$m$^{-3}$
and $B \gtrsim 1\,$T, indicating that the effect of bremsstrahlung
radiation reaction is generally negligible. It will typically impose an energy
limit an order of magnitude larger than that at which the synchrotron
reaction force takes effect. Note, however, that \Eq{eq:p0S} for the
location of the distribution maximum caused by synchrotron radiation
is a lower limit, accurate only to within a factor $\sim 5$ when
compared to numerical results. This acts in favor of the importance
of bremsstrahlung, so that a simple rule of thumb for the importance
of bremsstrahlung effects (in place of Eq.~\ref{eq:n20lim}) is
\begin{equation}
n_e[10^{20}\,\text{m}^{-3}] \gtrsim B[T]^2.
\end{equation}

It should be noted that in partially ionized plasmas containing high
atomic number impurities, such as those produced during disruption
mitigation scenarios with Massive Gas Injection in tokamaks, the
bremsstrahlung will be enhanced compared to the expressions given here
due to highly energetic electrons penetrating the charged cloud of
electrons around the atomic nuclei. This effect is described by an
atomic form factor which multiplies the cross-section
formulas~\cite{kochmotz1959}. However, the same is true for elastic
Coulomb collisions, as described in Ref.~\onlinecite{zhogolev},
leading to enhanced pitch-angle scattering with the consequence that
the synchrotron force experiences a similar increase in efficacy. The
enhanced scattering is thus not expected to significantly shift the
balance between the bremsstrahlung and synchrotron forces. A full
treatment of these effects, however, requires a detailed description
of the atomic physics involved and is beyond the scope of this paper.

\paragraph*{Curvature-induced synchrotron losses}--- 
It should be pointed out that, while the above is valid for straight
field-line geometry, additional effects come into play in toroidal
geometry.  An additional energy loss through synchrotron radiation
caused by the acceleration of the guiding center due to the
toroidicity adds a frictional term similar to the bremsstrahlung
radiation friction. In the high-energy limit, as derived in
Ref.~\onlinecite{andersson2001}, this orbit-induced synchrotron
radiation term takes the form
\begin{align}
\left(\frac{\rd p}{\rd t}\right)\sub{orbit} = -\frac{m_e c}{\tau_S}\frac{\rho_0^2}{R^2}\left(\frac{p_\parallel}{m_e c}\right)^4,
\end{align}
where $\rho_0 = m_e c /e B$ and $R$ is the average radius of curvature
of the magnetic field.  
To determine when this effect is important compared to the bremsstrahlung
losses, we may investigate at what electron energy the two reaction
forces become equal.
Assuming that, for a runaway electron, it is
valid to replace $p_\parallel \approx p \approx \gamma m_e c$, 
setting the loss terms as equal yields the condition
\begin{align}
\frac{1}{\tau_B}\gamma\ln 2\gamma = \frac{1}{\tau_S}\frac{\rho_0^2}{R^2}\gamma^4,
\end{align}
where we have assumed $\gamma \gg 1$. Inserting numerical values for
natural constants, we find 
\begin{align}
\frac{\gamma^3}{\ln 2\gamma} \approx  (1+Z\sub{eff}) 1.2\cdot 10^{4}n_e[10^{20}\,\text{m}^{-3}] R[\text{m}]^2.
\end{align}
Assuming a characteristic $\ln 2 \gamma = \ln 200 \approx 5.3$ -- corresponding
to runaway electron energies of approximately 50\,MeV -- we find 
that bremsstrahlung losses dominate orbit-induced synchrotron losses 
for relativistic factors smaller than
\begin{align}
\gamma \approx 40 \biggr(  (1+Z\sub{eff}) n_e[10^{20}\,\text{m}^{-3}] R[\text{m}]^2 \biggr)^{1/3}.
\label{eq:orbit condition}
\end{align}
As we have previously indicated, bremsstrahlung radiation loss will be
insignificant compared to the gyromotion-induced synchrotron radiation
loss unless $n_e[10^{20}\,\text{m}^{-3}] \gtrsim B[T]^2$.  In tokamak
plasmas this will primarily be the case during massive gas injection
scenarios. Assuming a hypothetical ITER massive gas injection with
$Z\sub{eff} = 10$, $n_e = 10^{21}\,$m$^{-3}$ and $R = 5\,$m, we find
that bremsstrahlung radiation loss will be dominant (as compared to
the orbit synchrotron radiation loss) for all electrons with
$\gamma \lesssim 560$, or energy lower than approximately 300 MeV. As
a consequence, in typical scenarios of interest the orbit-induced
synchrotron radiation loss will not affect our conclusions, although
some care must be taken to ensure that expected runaway-electron energies are
far from the threshold energy indicated by Eq.~(\ref{eq:orbit condition}).

We may note that, unlike the synchrotron radiation caused by the
gyromotion which is sensitive to pitch-angle scattering, the
orbit-induced radiation loss tends to be less significant in
comparison to bremsstrahlung radiation loss as the plasma effective
charge increases. Its effect is also independent of magnetic field
strength (cancelation occurring between $\tau\sub{S}$ and $\rho_0$),
and thus only depends on particle energy and radius of curvature.


\subsection{Numerical results}

As discussed in Section \ref{sec:Effect_on_dist}, bremsstrahlung effects 
can significantly reduce the average runaway energy under those
circumstances where it leads to a bump at lower particle
energies than when considering only synchrotron radiation
reaction. This shift in bump-on-tail location is one of the indicators of the importance of
bremsstrahlung effects, and to quantify it we use \textsc{CODE} to
study time-asymptotic electron distributions, for a variety of
plasma parameters. 
Although the (quasi-)steady-state distribution function is not always of practical
relevance in applications because of the time scales involved,
this method allows us to quantify the
relative importance of synchrotron and bremsstrahlung effects in a
robust way.
  
Numerical distributions have been calculated for a variety of
parameter sets, with parameters in the ranges:
$n_e \in [3\cdot 10^{18},1\cdot 10^{22}]$~m$^{-3}$, $B \in [0,4]$~T,
$\zeff \in [1,5]$ and $E/E_c \in [1.5,5]$, with and without bremsstrahlung
radiation reaction included, for $T_e=10$~keV. The results were
analyzed after $5\cdot 10^5$ thermal collision times, giving the
distributions ample time to reach steady state.

\begin{figure}
	\begin{center}
		\includegraphics[width=1\columnwidth, trim={0.35cm 0 0.9cm 0cm},clip]{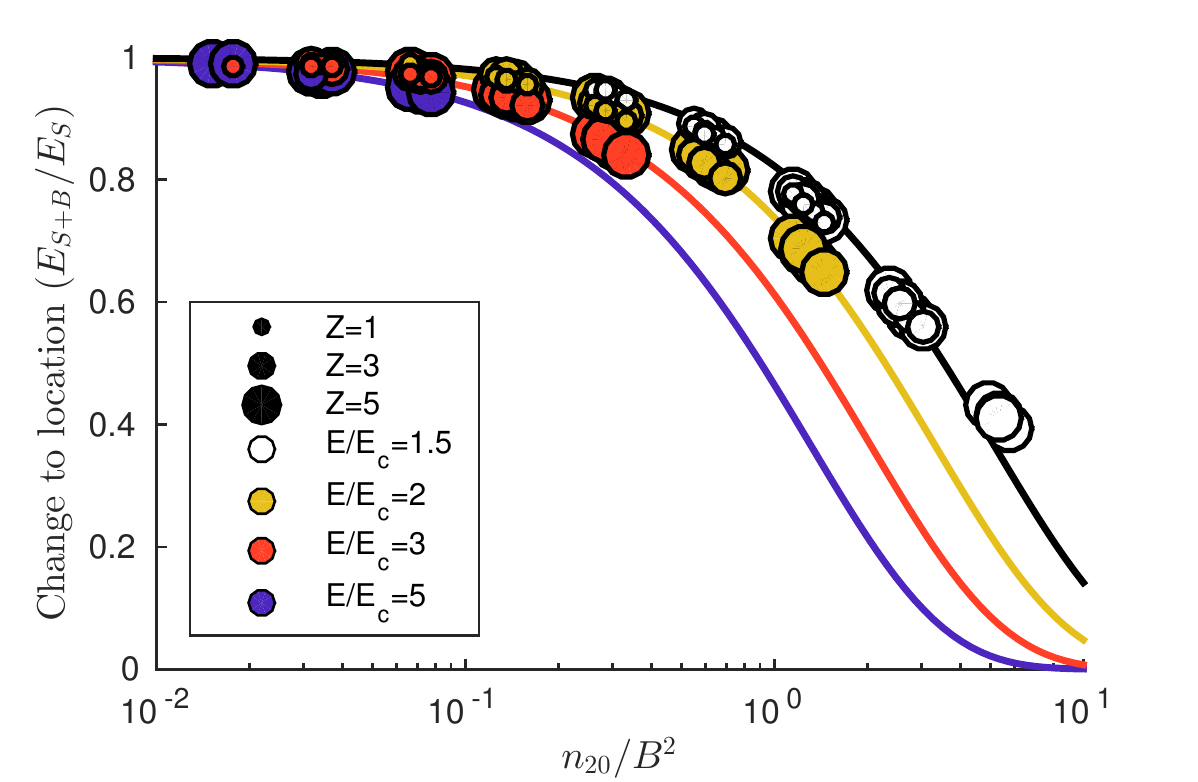}
		\caption{\label{fig:bump_location_energy_spread_scan}
                  Comparison between the location of the bump at
                  $p_\perp=0$ with and without bremsstrahlung
                  radiation reaction effects included, as a function of 
                  the ratio of electron density and magnetic field squared. 
                  Circles denote \textsc{CODE} results, with $\zeff$ indicated by the
                  size of the marker and $E/E_c$ by the color. The
                  solid lines are simple exponential fits to the data
                  points, provided for visual guidance.}
	\end{center}
\end{figure}

\begin{figure}[t]
	\begin{center}
		\includegraphics[width=1\columnwidth, trim={0.35cm 0 0.9cm 0cm},clip]{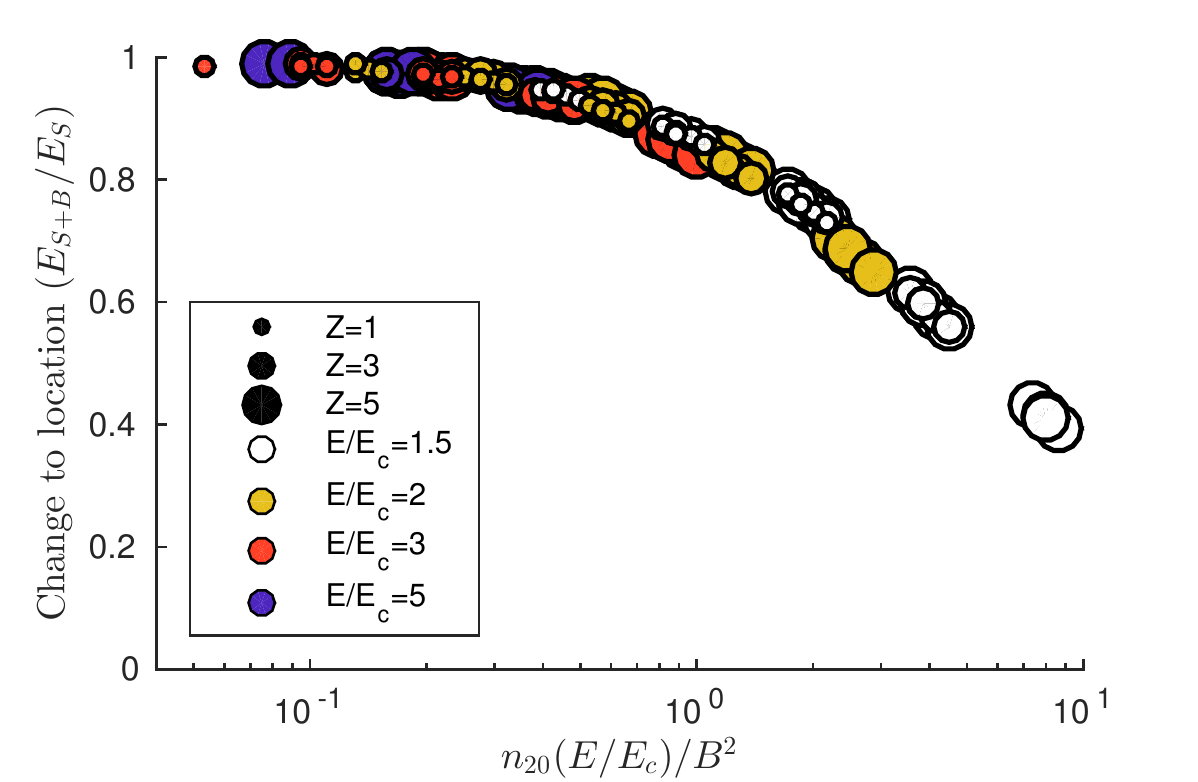}
		\caption{\label{fig:bump_location_energy_spread_EEc}  The same as \Fig{fig:bump_location_energy_spread_scan}, except that the x-axis is weighted by an additional factor $E/E_c$.}
	\end{center}
\end{figure}

Figure~\ref{fig:bump_location_energy_spread_scan} shows a comparison of
the location of the bump feature, with and without bremsstrahlung
effects. Only data points where the entire bump was contained on the
numerical grid in both the synchrotron-only and
synchrotron-and-bremsstrahlung cases are included, and bumps located
at energies below 9 MeV were omitted. The figure shows that the bump
appears at lower energies  as the density increases, with the change becoming
significant for $n_e[10^{20}\,\text{m}^{-3}]/B[T]^2$ of order unity. This is in good agreement
with the rule-of-thumb discussed in
Section~\ref{sec:Estimate_BS_vs_B}. There is also no
discernible dependence of the results on $\zeff$, although the
magnitude of the effect at a given $n_e/B^2$ increases with
increasing $E/E_c$. While this dependence on the electric field is not 
explicitly described by the formulas derived in the previous section, 
it is not inconsistent with our treatment. This is because Eq.~(\ref{eq:p0S}) 
describes only an approximate lower limit, correct within an
order-unity coefficient that can depend on the other parameters. By
numerically investigating the parametric dependence on the electric field strength,
we find that a more accurate condition for the importance of bremsstrahlung effects is given by
\begin{equation}
\frac{E}{E_c} n_e[10^{20}\,\text{m}^{-3}] \gtrsim B[T]^2,
\end{equation}
as illustrated in
\Fig{fig:bump_location_energy_spread_EEc}.

\begin{figure*}[htbp]
	\begin{center}
		\includegraphics[width=0.85\textwidth, trim={1.6cm 0 1.95cm 0.5cm},clip]{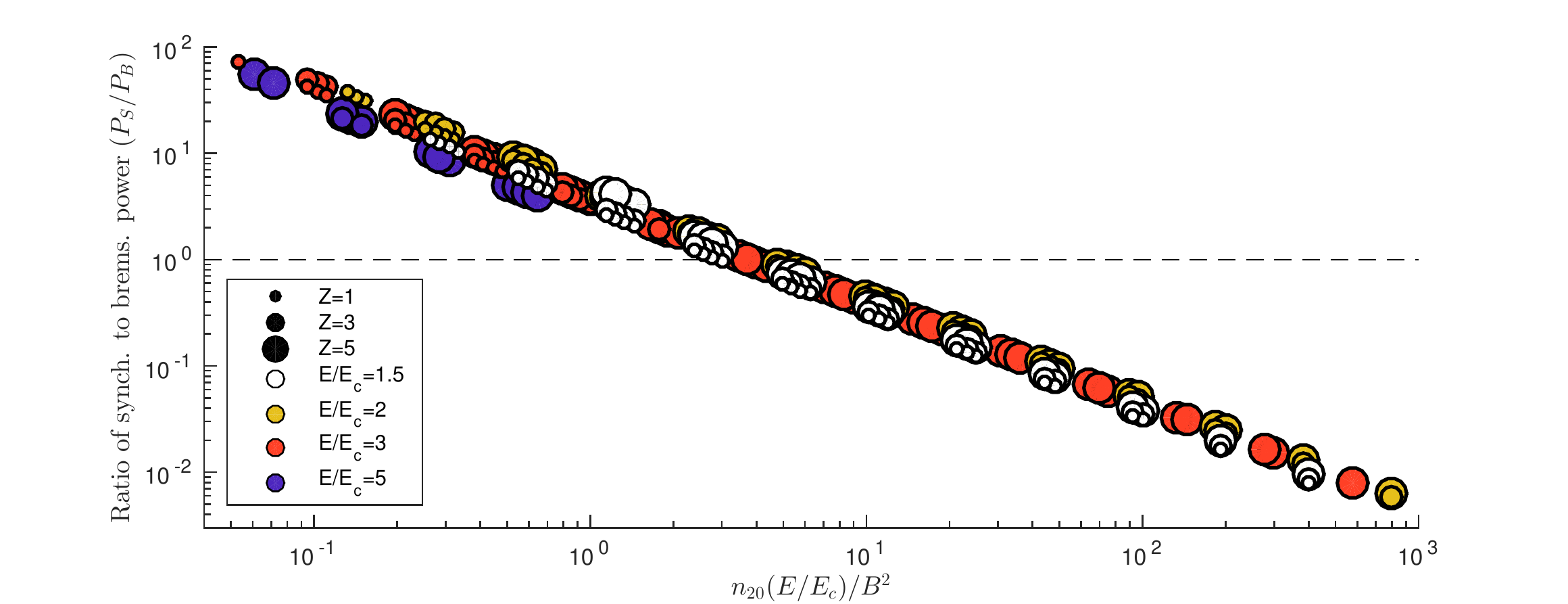}
		\caption{\label{fig:S_vs_B_power} Ratio of synchrotron to bremsstrahlung radiated power from distributions obtained with both synchrotron and bremsstrahlung effects included. $\zeff$ is indicated by the size of the marker and $E/E_c$ by its color. 
}
	\end{center}
\end{figure*}

Another indicator of the relative importance of synchrotron and
bremsstrahlung effects is the ratio of the power radiated by the two
processes, as the process taking away the most energy is likely to
have the largest impact on the distribution
function. Figure~\ref{fig:S_vs_B_power} shows this ratio, in
distributions obtained with both synchrotron and bremsstrahlung
effects included. The emitted powers are calculated as the energy
moments of the respective terms in the kinetic equation. We note that
for $(E/E_c) n_e[10^{20}\,\text{m}^{-3}]/B[T]^2\lesssim 4$, the radiated power is predominantly
synchrotron radiation, whereas when $(E/E_c) n_e[10^{20}\,\text{m}^{-3}]/B[T]^2\gtrsim 4$,
bremsstrahlung is dominant. This reaffirms the conclusion drawn from
\Fig{fig:bump_location_energy_spread_EEc} that bremsstrahlung starts
to have a significant impact on the distribution at around
$(E/E_c) n_e[10^{20}\,\text{m}^{-3}]/B[T]^2\sim 1$.  Again, there is no discernible dependence on
$\zeff$.

\begin{figure}
	\begin{center}
		\includegraphics[width=1\columnwidth, trim={0.3cm 0.15cm 0.4cm 0.35cm},clip]{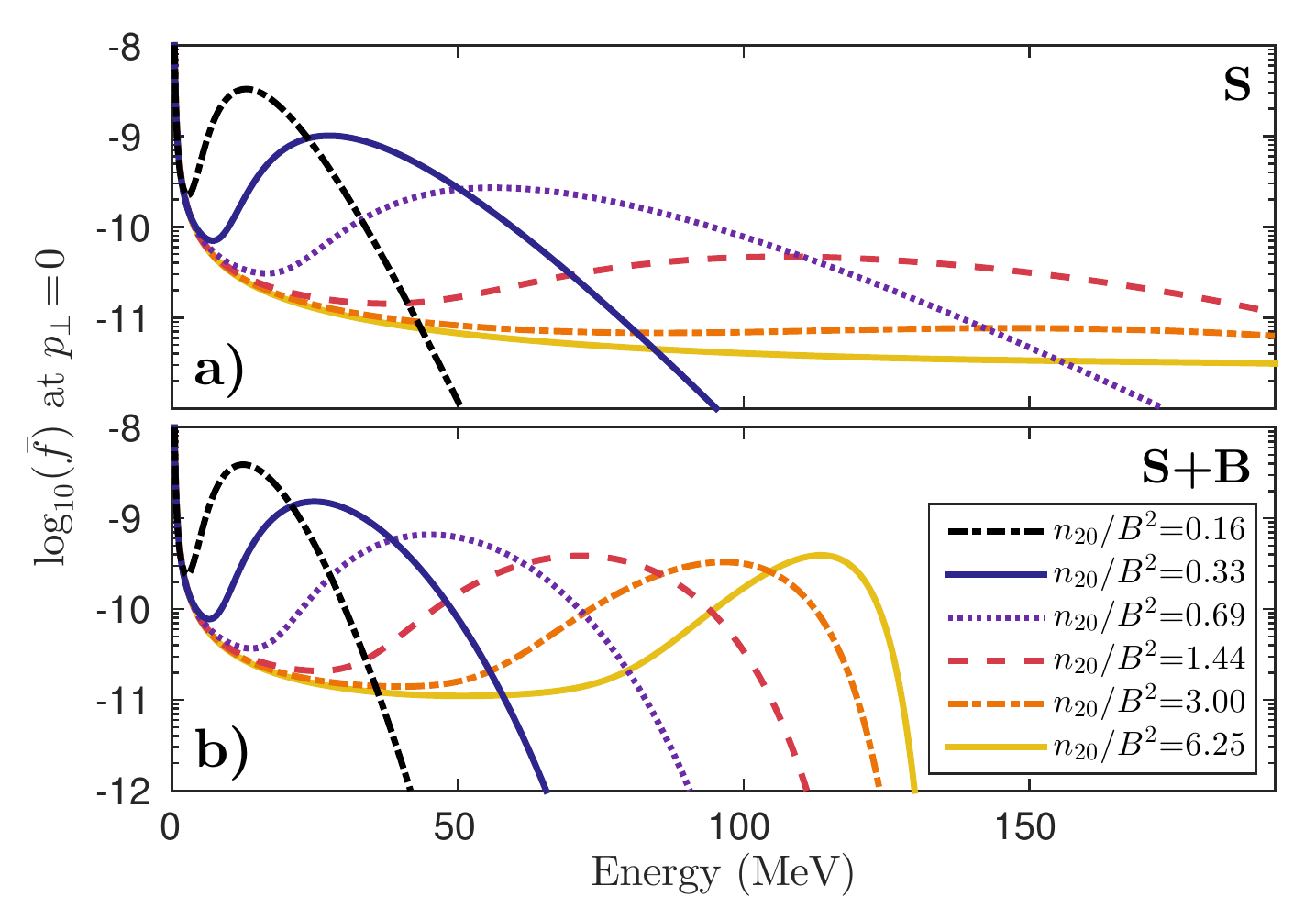}
		\caption{\label{fig:dists_nB2} Distributions at $p_\perp=0$ for several values of $n_e$, with a) synchrotron and b) synchrotron and bremsstrahlung effects included. The parameters are $T_e=10$~keV, $\zeff=3$, $E/E_c=2$ and $B=4$~T.}
	\end{center}
\end{figure}

Looking directly at the distribution function can also provide important
insight. Comparing the distributions in \Fig{fig:dists_nB2}b
(synchrotron and bremsstrahlung) to those in \Fig{fig:dists_nB2}a
(synchrotron only), it is clear that for the lowest densities shown,
the distributions are very similar. The effect on the distribution is
thus synchrotron-dominated in this case. For higher densities, the
distributions do however become increasingly different, and above
$ n_e[10^{20}\,\text{m}^{-3}]/B[T]^2\!\sim\!1$ they are distinctly bremsstrahlung-dominated, as
would be expected.  Several different measures thus seem to agree that
bremsstrahlung has an important effect on the fast electron
distribution function for densities such that
\mbox{$ n_e[10^{20}\,\text{m}^{-3}]\gtrsim B[T]^2$}.

\section{Conclusions}
\label{sec:conclusions}

The effect of bremsstrahlung on the runaway electron distribution was
investigated by extending the 2D Fokker-Planck equation solver \textsc{CODE}
with a simplified description of the bremsstrahlung process, allowing
rapid calculations. This method provides qualitative insight into the
effect of bremsstrahlung on the distribution function, and allows for
quantitative predictions of the maximum energy attainable by
runaway electrons.  In particular we have investigated the effect of
bremsstrahlung radiation losses in combination with synchrotron
radiation losses. 

We show that the inclusion of bremsstrahlung losses leads to non-monotonic
(``bump'') features in the electron distribution function. Similar
features have previously been described in the presence of synchrotron
radiation, but bremsstrahlung leads to much stronger gradients in
the magnitude of the momentum, as well as more significant spreading in
pitch-angle.  The combined effect of synchrotron and bremsstrahlung
therefore leads to a bump that limits the particle energy more
efficiently than in the case of pure synchrotron radiation reaction. 
From force balance considerations we derived an expression
for the maximum electron energy, which shows excellent agreement with
numerical simulations using \textsc{CODE}.

We show that if the condition $(E/E_c) n_e[10^{20}\,$m$^{-3}] \gtrsim B[T]^2$
is fulfilled, the effect of bremsstrahlung radiative losses on the electron
distribution function are expected to be more important than
synchrotron radiative losses. The importance of bremsstrahlung
compared to synchrotron was found to be insensitive to plasma
composition due to the similar parametric dependence on ion species of 
the bremsstrahlung reaction rate and pitch-angle scattering (related to the efficacy of synchrotron
losses). This means that for typical tokamak-runaway scenarios we expect synchrotron radiation
losses to dominate. In an ITER-like plasma with $B\simeq 5$\,T, for example, 
the required electron density for significant stopping
power due to bremsstrahlung (as compared to synchrotron losses) is of
order $n_e = 3\cdot 10^{21}$\,m$^{-3}$, which is significantly larger
than the expected operational density of order
$n_e = 1\cdot10^{20}$\,m$^{-3}$. However, in the case of massive gas
injection for disruption mitigation, density increases of this order
are possible, and in such scenarios it will be important to account
for the bremsstrahlung losses, as well as synchrotron radiation losses, when modeling the slowing-down of 
fast electrons.

Note, finally, the limitations of the present study; we have considered
steady-state solutions of the distribution function at low electric
fields ($E/E_c$ of order unity), where particle fluxes in momentum
space are in balance. In transient processes, such as the initial
slowing-down following a spike in the electric field strength, or other acceleration
process, it is conceivable that the balance between bremsstrahlung and
synchrotron losses is temporarily shifted before quasi-equilibrium is
established. This typically happens on a time-scale of a few
(relativistic) collision times
$\tau_c = 4\pi\varepsilon_0^2 m_e^2 c^3/(e^4 n_e \ln\Lambda ) \approx
0.3 / (\ln\Lambda n_e[10^{20}$m$^{-3}])$\,s.
Furthermore, as we have treated bremsstrahlung as a continuous
process, the details of the shape of the distribution functions as
shown in Figs.~\ref{fig:dists_S_B_SB} and \ref{fig:dists_nB2} should be considered 
only indicative. A more
sophisticated treatment accounting for the finite photon energies is likely to
lead to a broader bump on the runaway tail due to the diffusive nature of the resulting electron motion. However, we expect the
qualitative trends (bump-on-tail formation, approximate location of
the local maximum) to be well represented by the model adopted
here. Finally, we have restricted this study to consider only fully
ionized plasmas. In post-disruption plasmas with runaway mitigation by
massive gas injection of high-$Z$ atoms, where temperatures are often
below $10\,$eV, both the bremsstrahlung and elastic scattering
cross-sections are modified to account for the presence of
bound electrons, due to incomplete screening of the charged nucleus at
large incident energies. These effects would act to increase both the
synchrotron (because of enhanced pitch-angle scattering) and
bremsstrahlung losses compared to the predictions of our model, and a
more detailed description of the atomic physics involved would be
required for a more accurate simulation of such scenarios.


\section*{\normalsize {\sffamily ACKNOWLEDGEMENTS}}

\noindent 
This work has been carried out within the framework of the EUROfusion Consortium and has received funding from the Euratom research and training programme 2014-2018 under grant agreement No 633053.
The views and opinions expressed herein do not necessarily reflect those of the European Commission.
This project has received funding from the Knut and Alice Wallenberg
Foundation and Vetenskapsr\aa det. 


\end{document}